\documentclass[11pt]{article}
\usepackage{epsfig,eucal}

\usepackage{graphicx,epsfig}

 \usepackage{amsfonts,amssymb}
\usepackage{amsmath}
\usepackage{youngtab}

\usepackage{amsthm}
\theoremstyle{plain}
\newtheorem{thm}{Theorem}

\theoremstyle{definition}

\theoremstyle{remark}

\usepackage{bm}
\usepackage{color}
\definecolor{dark-green}{rgb}{0,0.7,0}
\definecolor{dark-blue}{rgb}{0,0.2,0.5}
\definecolor{med-blue}{rgb}{0,0.7,1}
\definecolor{mblue}{rgb}{0,0.2,1}
\definecolor{cnc}{rgb}{0.8,0,0}
\definecolor{light-red}{rgb}{1,0.8,0.8}
\definecolor{dark-yellow}{rgb}{1,0.8,0}
\definecolor{light-blue}{rgb}{0.8,0.9,1}
\definecolor{verylight-blue}{rgb}{0.93,0.95,1}
\definecolor{light-yellow}{rgb}{1,0.9,0.8}
\definecolor{grey}{gray}{0.88}
\definecolor{new-green}{rgb}{0.5,0.5,0.6}

\def\a{\alpha} \def\b{\beta} \def\g{\gamma}

 \def\S{{\cal{S}}} \def\A{{\cal{A}}}

\begin{document}

\title{Quadratic invariants of the elasticity tensor}
\author{Yakov Itin\\
  Inst. Mathematics, Hebrew Univ. of Jerusalem, \\
  Givat Ram,
  Jerusalem 91904, Israel,\\
  and Jerusalem College of Technology,
  Jerusalem 91160, Israel,\\
  email: itin@math.huji.ac.il\\ \\
}

\date{\today, {\it file Quadratic6.tex}}
\maketitle
\begin{abstract} 
We study the quadratic invariants of the
  elasticity tensor in the framework of its unique irreducible
  decomposition.  The key point is that this 
  decomposition generates the direct sum
  reduction of the elasticity tensor space. The
  corresponding subspaces are completely independent and even
  orthogonal relative to the Euclidean (Frobenius) scalar product. We
  construct a basis set of seven quadratic invariants that emerge in
  a natural and systematic way. Moreover, the completeness of this basis and
 the independence of the basis tensors follow immediately from the direct sum
  representation of the elasticity tensor space.
 We define the Cauchy  factor of an anisotropic material as  a dimensionless measure of a closeness to a pure Cauchy material and a similar isotropic factor is as a measure for a closeness of an anisotropic material to its isotropic prototype. For  cubic crystals, these factors are explicitly displayed and cubic crystal average of an arbitrary elastic material is derived.
\end{abstract}

{\it {Key index words:} {anisotropic} elasticity tensor,
  irreducible decomposition, quadratic invariants} 

\pagestyle{myheadings} \markboth{Y. Itinl} {Y. Itin \hspace{180pt} Quadratic invariants of the elasticity tensor}

\maketitle
\section{Introduction }
In the linear elasticity theory of anisotropic materials, the relation between the strain tensor $u_{ij}$ and the stress tensor $\sigma^{ij}$, the generalized Hooke's law,  is expressed by the use of the elasticity (stiffness) tensor $C^{ijkl}$
\begin{equation}\label{intro1}
\sigma^{ij}=C^{ijkl}u_{kl}\,.
\end{equation}
In 3-dimensional space, a generic 4th order tensor has  81 independent components. However, due to the standard symmetry assumptions for the stress and strain tensors, 
\begin{equation}\label{intro2}
C^{ijkl}=C^{jijkl}=C^{klij}\,,
\end{equation}
the elasticity tensor is left with 21 independent components only. 
These components are not really the intrinsic characteristics of the material  because they depend on the choice of the coordinate system. Thus, in order to deal with the proper material parameters, one must look for the {\it invariants of the elasticity tensor}. 

There are only two linearly independent  invariants of the first order of $ C^{ijkl}$.  They  are usually taken as follows, see \cite{Norris},
 \begin{equation}\label{intro2a}
A_1= C^{ij}{}_{ij}\stackrel{*}= C_{ijij} \qquad {\rm and} \qquad A_2=C_i{}^{ij}{}_j\stackrel{*}= C_{iijj}\,.
\end{equation}
Here and subsequently, we use the standard tensor conventions that strictly distinguish between covariant and contravariant indices. In these notations,  two repeated indices can appear only in up-down positions and summation  for only two such   repeated indices is assumed. 
The indices of a tensor can be raised/lowered by the use of the metric tensor  $g^{ij}/g_{kl}$. For instance, the lower components of the elasticity tensor are defined as $C_{ijkl}=g_{ii'}g_{jj'}g_{kk'}g_{ll'}C^{i'j'k'l'}$. 
Since in the elasticity literature a simplified notation  is frequently used, we provide both notations in most cases  and relate the corresponding quantities by the sign $\stackrel{*}=$ , as in (\ref{intro2}). Notice that this shorthand  notation is applicable only in Euclidean space endowed with rectangular coordinates. 

Quadratic invariants of the elasticity tensor were studied by Ting \cite{Ting}. He presented two such invariants, 
 \begin{equation}\label{intro3}
B_1=C^{ijkl}C_{ijkl}\stackrel{*}= C_{ijkl}C_{ijkl} \qquad {\rm and} \qquad B_2=C_i{}^{ikl}C^j{}_{jkl}\stackrel{*}= C_{iikl}C_{jjkl}\,.
\end{equation}
Ahmad \cite{Ahmad} has  contributed the two additional quadratic invariants,  
 \begin{equation}\label{intro4}
B_3=C_i{}^{ikl}C^j{}_{kjl}\stackrel{*}= C_{iikl}C_{jkjl} \qquad {\rm and} \qquad B_4=C_i{}^{kli}C^j{}_{klj}\stackrel{*}= C_{kiil}C_{kjjl}\,.
\end{equation}
He also proved that the set of seven quadratic invariants
\begin{equation}\label{intro5}
\bigg{\{}A_1^2\,,\quad A_2^2\,,\quad A_1A_2\,,\quad B_1\,,\quad B_2 \,,\quad B_3\,, \quad B_4\bigg{\}}
\end{equation}
is linearly independent. Norris \cite{Norris}  studied the problem of the quadratic invariants  and proved that the set (\ref{intro5}) is complete. It means that every quadratic invariant of the elasticity tensor is a linear combination of the seven invariants listed in (\ref{intro5}). In order to prove this fact, Norris presented a generic quadratic invariant   in the form
\begin{equation}\label{intro6}
I=f_{ijklpqrs}C^{ijkl}C^{pqrs}\,,
\end{equation}
where $f_{ijklpqrs}$ is a numerical tensor and provided a detailed analysis of this tensor. It is proven that due to the symmetries (\ref{intro2}), the  most components of $f_{ijklpqrs}$ vanish and a lot of the remaining components are linearly related. This way, exactly seven invariants are left over. 
In particular, Norris demonstrated that a rather natural additional invariant 
\begin{equation}\label{intro7}
B_5=C^{ijkl}C_{ikjl}\stackrel{*}= C_{ijkl}C_{ikjl}
\end{equation}
can be in fact represented as a linear combination of the invariants listed in (\ref{intro5}), namely 
\begin{equation}\label{intro8}
B_5=\frac 12 A_1^2+\frac 12 A_2^2-A_1A_2+B_1-2B_2+4B_3-2B_4\,.
\end{equation}

In this situation, some principal questions arise: 
\begin{itemize}
\item Is there  some preferable basis of  quadratic invariants?
\item Is there  a systematic way to construct such a basis?
\item Which quadratic invariants can be used as  characteristic parameters  for  different elastic materials?
\item Which physical interpretation can be given to various quadratic invariants of the elasticity tensor?
\end{itemize}
In the current paper, we analyze the quadratic invariants problem in the framework of the unique irreducible decomposition of the elasticity tensor. In Section 2, we present the principle algebraic facts of this decomposition. The key point that such resolution  of the elasticity tensor in the simple pieces generates the  direct sum reduction of the elasticity tensor space. The corresponding subspaces are completely independent and even orthogonal relative to the Euclidean (Frobenius) scalar product. In this framework, we construct in Section 3 the basis set of seven quadratic invariants that emerge in a natural and systematic way. Moreover the completeness of this basis and its independence of basis tensors follow  immediately from the direct sum representation of the elasticity tensor space. We compare this basis to the basis given in (\ref{intro5}). In section 4, we provide some applications of quadratic invariants to physics motivated problems. We define Cauchy factor for an arbitrary anisotropic material. It can be used as a measure of deviation of a material from an analogical Cauchy material. We also prove Fedorov's relation for an isotropic  material closest to an anisotropic one. This result follows immediately from the irreducible orthogonal decomposition.  Correspondingly, we define  the isotropic factor that measures the closeness of an anisotropic material to its isotropic prototype. In section 5, we study a simplest non-trivial example of cubic crystal. It is naturally represented by three independent quadratic invariants. This low-dimensional case allows the visualization  of the Cauchy and isotropic factors. 
Corresponding graphs are presented. In Conclusion section we present our main  results and propose some possible direction for future investigations. 

\section{Irreducible decomposition}
To describe the irreducible decomposition of the elasticity tensor we first observe two groups acting on it simultaneously, see \cite{Itin-Hehl2}. For an arbitrary tensor of the range $p$ defined on ${\mathbb R}^n$, these are the permutation (symmetry) group $S_p$ and the group of rotations $SO(n,{\mathbb R})$. 
\subsection{Irreducible decomposition under the permutation group}
The symmetry group $S_4$ provides permutations of the elasticity tensor indices. 
The decomposition of $C^{ijkl}$ under this group is described by two Young diagrams: 
\begin{equation}\label{irr3}
  \Yvcentermath1\yng(1) \oplus \yng(1)\otimes\yng(1)\otimes\yng(1)
  =\yng(4)\oplus \yng(2,2)\,.
\end{equation}
All other 4-the order Young's diagrams are non-relevant in our case due to the original symmetries (\ref{intro2}) of $C^{ijkl}$.
The left-hand side of (\ref{irr3}) represents a generic 4th rank tensor.  On the
right-hand side, two diagrams  describe  two tensors of different symmetries. The explicit expression of these tensors can be  computed by the use of the corresponding symmetrization and antisymmetrization operations. For the elasticity tensor, the result is obvious.  
The first (row) diagram  represents the totally symmetric tensor. The second (square) diagram represents an additional tensor that can be considered merely as  a remainder. 

As a result, we arrive at the unique and  irreducible decomposition of the elasticity tensor under the action of  the group $S_4$:
\begin{equation}\label{decomp}
  C^{ijkl}=S^{ijkl}+A^{ijkl}\,,\end{equation}
with 
\begin{equation}\label{firsty}
  S^{ijkl}:=C^{(ijkl)}=\frac 13\left(C^{ijkl}+C^{iklj}+ C^{iljk}\right)\,
\end{equation}
and 
\begin{equation}\label{decomp1}
  A^{ijkl}:=C^{ijkl}-C^{(ijkl)}=\frac 13\left(2C^{ijkl}-C^{ilkj}-C^{iklj}\right)\,.
\end{equation}
We use the standard normalized Bach parentheses for symmetrization and antisymmetrization of indices. We observe that every antisymmetrization of $ S^{ijkl}$ gives zero and every symmetrization preserves this tensor. As for the second part $A^{ijkl}$, its total symmetrization $A^{(ijkl)}$ vanishes. Consequently, we have a useful symmetry relation
\begin{equation}\label{decomp1a}
  A^{ijkl}+A^{iklj}+A^{iljk}=0\,.
\end{equation}

As it is shown in \cite{Cauchy} and \cite{Itin-Hehl}, the equation $A^{ijkl}=0$ describes the well-known Cauchy relation. Thus, we call $S^{ijkl}$ the {\it Cauchy part} and $A^{ijkl}$ the {\it non-Cauchy part} of the elasticity tensor .

Observe the main algebraic properties of this decomposition:
 \begin{itemize}
\item The partial tensors $S^{ijkl}$ and $A^{ijkl}$ satisfy the minor symmetries,
 \begin{equation}\label{alg1}
S^{[ij]kl}=S^{ij[kl]}=0\qquad {\rm and}\qquad  A^{[ij]kl}=A^{ij[kl]}=0\,,
\end{equation}
and the major symmetry,
 \begin{equation}\label{alg1a}
S^{ijkl}=S^{klij}\qquad {\rm and}\qquad  A^{ijkl}=A^{klij}\,.
\end{equation}
Thus, these partial tensors can themselves serve as elasticities of some hypothetic material. 
\item Moreover, any additional symmetrization or antisymmetrization preserves the tensors $S^{ijkl}$ and $A^{ijkl}$ or nullifies them. 
\item The decomposition (\ref{decomp}) is preserved under arbitrary linear transformations. Thus, it can be referred to as irreducible $GL(3,{\mathbb R})$-decomposition.
\item The irreducible decomposition of the tensor $C$
provides the decomposition of the corresponding tensor space ${\cal C}$ into a {\it direct
sum} of two subspaces ${\cal S}\subset{\cal C}$ (for the tensor $S$)
and ${\cal A}\subset{\cal C}$ (for the tensor $A$),
\begin{equation}\label{alg1b}
{\cal C}={\cal S}\oplus{\cal A}\,.
\end{equation}
 In particular, we have
\begin{equation}\label{alg1c} 
{\rm dim} \, {\cal C}=21\,,\qquad {\rm dim}
    \, {\cal S}=15\,,\qquad {\rm dim} \, {\cal A}=6\,.\end{equation}
\item The irreducible pieces $S^{ijkl}$ and $A^{ijkl}$ are orthogonal to one  another in the following sense: 
\begin{equation}\label{ORT1}
S^{ijkl}A_{ijkl}=0\,.
\end{equation}
Indeed, 
\begin{eqnarray}\label{rest**A}
S^{ijkl}A_{ijkl}&=&S^{(ijkl)}A_{ijkl}=S^{(ijkl)}A_{(ijkl)}=0 \,.
\end{eqnarray}
\item ``Pythagorean theorem:"  The Euclidean (Frobenius)  squares of the tensors 
\begin{equation}\label{ORT2}
{\widetilde C}^2=C^{ijkl}C_{ijkl}\,,\qquad {\widetilde S}^2=S^{ijkl}S_{ijkl}\,,\qquad {\widetilde A}^2=A^{ijkl}A_{ijkl}
\end{equation}
 satisfy the relation 
\begin{eqnarray}\label{ORT3}
{\widetilde C}^2={\widetilde S}^2+{\widetilde A}^2 \,.
\end{eqnarray}
\end{itemize}
\subsection{Irreducible decomposition under the rotation group}
We are looking now for the decomposition of the elasticity tensor under the action of the group of rotations.  The following fact is due to the classical   theory of invariants: Relative to the subgroup $SO(3)$ of $GL(3)$, a basis of an arbitrary system of tensors coincides with a basis of the same system with the metric tensor added \cite{Weyl}. We use the Euclidean metric tensor $g_{ij}$. In rectangular coordinates,  $g_{ij}={\rm diag}(1,1,1)$.  

We start with the totally symmetric Cauchy part $S^{ijkl}$. 
From the contraction of  $S^{ijkl}$ with the metric tensor, we construct a unique symmetric second-rank tensor 
 \begin{equation}\label{S2}
   S^{ij}:=g_{kl}S^{ijkl}\stackrel{*}=S_{ijkk}=\frac 13 \left(C_{ijkk}+2C_{ikkj}\right)\,
\end{equation}
and a unique scalar 
\begin{equation}\label{S2a}
  S:=g_{ij}S^{ij}\stackrel{*}=S_{iikk}=\frac 13 \left(C_{iikk}+2C_{ikki}\right)\,.
\end{equation}
We denote the traceless part of the tensor $S^{ij}$ as
\begin{equation}\label{S00} P^{ij}:=S^{ij}-\frac 13
  Sg^{ij}\,,\qquad\text{with}\qquad g_{ij}P^{ij}=0\,.
\end{equation}
Now we turn to the decomposition of the  tensor $S^{ijkl}$. We denote the two subtensors
 \begin{equation}\label{sub1}
{}^{(1)}\!S^{ijkl}:=\a Sg^{(ij}g^{kl)}\,,\qquad {}^{(2)}\!S^{ijkl}:=\b P^{(ij}g^{kl)}\,.
\end{equation}
It can be checked now by the straightforward calculations that the remainder
\begin{equation}\label{sub3}
  R^{ijkl}:=S^{ijkl}-{}^{(1)}\!S^{ijkl}-{}^{(2)}\!S^{ijkl}\,
\end{equation}
is totally traceless if and only if
 \begin{equation}\label{sub3a}
\a= \frac 15\,,\qquad \b=\frac 67\,.
\end{equation}

Hence, we obtain the decomposition of the totally symmetric tensor $S^{ijkl}$  into  the sum of three independent pieces:
\begin{equation}\label{sub5} 
S^{ijkl}={}^{(1)}\!S^{ijkl}+{}^{(2)}\!S^{ijkl}+{}^{(3)}\!S^{ijkl}\,,
\end{equation}
where
 \begin{equation}\label{sub2}
  {}^{(1)}\!S^{ijkl}=\frac 15 Sg^{(ij}g^{kl)}=\frac 1{15} S\left(g^{ij}g^{kl}+g^{ik}g^{jl}+g^{il}g^{jk}\right)\,.
\end{equation}
\begin{equation}\label{sub1}
   {}^{(2)}\!S^{ijkl}=
\frac 67 P^{(ij}g^{kl)}=\frac 17 \left(P^{ij}g^{kl}+P^{ik}g^{jl}+P^{il}g^{jk}+P^{jk}g^{il}
+P^{jl}g^{ik}+P^{kl}g^{ij}\right)\,.
\end{equation}
and 
\begin{equation}\label{sub1x}
   {}^{(3)}\!S^{ijkl}=R^{ijkl}\,.
\end{equation}
These pieces are unique and invariant under the action of the group $SO(3)$.   Moreover, the corresponding  subspaces ${}^{(1)}\!{\cal S}$, ${}^{(2)}\!{\cal S}$ and ${}^{(3)}\!{\cal S}$ are mutually orthogonal. Indeed, for $A, B=1,2,3$ with $A\ne B$, 
\begin{equation}\label{ort} 
{}^{(A)}\!S_{ijkl}\,{}^{(B)}\!S^{ijkl}=0\,.
\end{equation}
This fact follows immediately from the tracelessness of the tensors $P^{ij}$ and $R^{ijkl}$.  

Consequently, the vector space ${\cal S}$ of the totally symmetric tensor $S^{ijkl}$ is decomposed into the direct sum of three subspaces
\begin{equation}\label{decompS1} 
{\cal S}={}^{(1)}\!{\cal S}\oplus{}^{(2)}\!{\cal S}\oplus{}^{(3)}\!{\cal S}
\end{equation}
with the corresponding dimensions
\begin{equation}\label{decompS2} 
15=1\oplus 5\oplus 9\,.
\end{equation}

We turn now to the second part of the elasticity tensor. 
The irreducible piece $A^{ijkl}$  is a fourth rank tensor with 6 independent components.   
It is quite naturally that it can be represented as a symmetric second-rank tensor, see \cite{Backus},\cite{Haus}, \cite{Cauchy}, and \cite{Itin-Hehl} for  detailed discussions. 
We define 
\begin{equation}\label{Del}
  \Delta_{mn}:=\frac 13\epsilon_{mil}  \epsilon_{njk}A^{ijkl}\,,
\end{equation}
where $\epsilon_{ijk}=0,\pm 1$ denotes the 3-dimensional Levi-Civita permutation pseudo-tensor. 
Consequently, $\Delta_{mn}$ is a symmetric tensor, $\Delta_{mn}=\Delta_{nm}$, and we have 
\begin{equation}\label{Delta-1}
 A^{ijkl}=\epsilon^{im(k}\epsilon^{l)jn}\Delta_{mn}\,.
\end{equation}

For a proof of this proposition it is enough to substitute (\ref{Del}) into (\ref{Delta-1}) and to apply the standard relations for  the Levi-Civita  pseudo-tensor. 

In order to decompose the  non-Cauchy part $A^{ijkl}$, it is convenient
to use its representation by the tensor density $\Delta_{ij}$. 
We denote
\begin{equation}\label{sub6a}
 A:=g^{mn}\Delta_{mn}\,.
\end{equation} 
By using the relation 
\begin{equation}\label{sub6a1}
 g^{mn}\epsilon_{mil}\epsilon_{njk}=g_{ij}g_{lk}-g_{ik}g_{jl}\, 
\end{equation}
and Eq.(\ref{Delta-1}), we derive 
\begin{equation}\label{sub6a2}
A=\frac 1{3}\left(g_{ij}g_{lk}-g_{ik}g_{jl}\right)A^{ijkl}\stackrel{*}= \frac 1{3}\left(A_{iikk}-A_{ikik}\right)=\frac 1{3}\left(C_{iikk}-C_{ikik}\right)\,.
\end{equation}
The  tensor density  $\Delta_{ij}$ can be  decomposed into the scalar and  traceless pieces:
\begin{equation}\label{sub6}
  \Delta_{ij}=Q_{ij}+\frac 13A  g_{ij}\,,
\end{equation}
where the traceless piece is given by
\begin{equation}\label{sub6b}
Q_{ij}:=\Delta_{ij}-\frac 13 A g_{ij}\,.
\end{equation}
Substituting (\ref{sub6}) into (\ref{Delta-1}), we obtain 
\begin{equation}\label{sub7}
A^{ijkl}={}^{(1)}\!A^{ijkl}+{}^{(2)}\!A^{ijkl}\,,
\end{equation}
where the scalar  part is given by 
\begin{equation}\label{sub8}
{}^{(1)}\!A^{ijkl}:=\frac 16
A\left(2g^{ij}g^{kl}-g^{il}g^{jk}-g^{ik}g^{jl}\right)\,
\end{equation}
and the remainder reads
\begin{equation}\label{sub8a}
{}^{(2)}\!A^{ijkl}:=\frac 12\left(\epsilon^{ikm}\epsilon^{jln}+
\epsilon^{ilm}\epsilon^{jkn}\right)Q_{mn} \,.
\end{equation}
We recall that the tensor $Q_{mn}$ is symmetric and traceless. 
Since the product of two Levi-Civita pseudo-tensors is represented by the determinant of the metric tensor $g^{ij}$, the latter equation can be rewritten as 
\begin{equation}\label{sub8aa}
{}^{(2)}\!A^{ijkl}:=g^{ik}Q^{jl}+g^{jk}Q^{il}+g^{il}Q^{jk}+
g^{jl}Q^{ik}-2g^{kl}Q^{ij}-2g^{ij}Q^{kl} \,.
\end{equation}
The decomposition given  in Eq.(\ref{sub7}) is unique, invariant, and irreducible under the action of the rotation group $SO(3,{\mathbb R})$ and of the permutation group $S_4$. 

Correspondingly, 
the vector space ${\cal A}$ of the tensor $A^{ijkl}$ is irreducibly decomposed into the direct sum of two subspaces 
\begin{equation}\label{sub9}
{\cal A}={}^{(1)}\!{\cal A}\oplus {}^{(2)}\!{\cal A}\,,
\end{equation}
with the corresponding dimensions
\begin{equation}\label{sub9}
6=1\oplus 5\,.
\end{equation}
Since the trace of ${}^{(2)}\!A^{ijkl}$ equals zero, these subspace are orthogonal to one  another,
\begin{equation}\label{sub9a}
{}^{(1)}\!A_{ijkl}{}^{(2)}\!A^{ijkl}=0\,.
\end{equation}
Collecting our  results, we formulate the following
\begin{thm} 
Under the simultaneous action of the groups $S_4$ and  $SO(3,{\mathbb R})$, the elasticity
tensor is uniquely irreducibly decomposed into the sum of five parts
\begin{equation}\label{sub10}
{C^{ijkl}=\sum^5_{A=1}{}^{(A)}C^{ijkl}=
 \left({}^{(1)}\!S^{ijkl}+{}^{(2)}\!S^{ijkl}+{}^{(3)}\!S^{ijkl}\right)+
\left({}^{(1)}\!A^{ijkl}+{}^{(2)}\!A^{ijkl}\right)}\,.
\end{equation}
This decomposition corresponds to the direct sum decomposition of the vector space of the elasticity tensor into five subspaces
\begin{equation}\label{ORT3}
{\cal C}= \left({}^{(1)}{\cal C}\oplus{}^{(2)}{\cal C}\oplus{}^{(3)}{\cal C}\right)\oplus \left({}^{(4)}{\cal C}\oplus{}^{(5)}{\cal C}\right)\,,
\end{equation}
with the dimensions
\begin{equation}\label{ORT3a}
21= \left(1\,\,\oplus\, \,5\,\,\oplus\, \,9\right) \,\,\oplus\,\left(1\,\,\oplus \,\,5\right)\,,
\end{equation}
The irreducible pieces are orthogonal to one  another: For $A\ne B$
\begin{equation}\label{ORT2}
{}^{(A)}C_{ijkl}{}^{(B)}C^{ijkl}=0\,.
\end{equation}
The Euclidean squares, $C^2=C_{ijkl}C^{ijkl}$ and ${}^{(A)}C^2={}^{(A)}C_{ijkl}\,{}^{(A)}C^{ijkl}$ with $A=1,\cdots,5$, fulfill the ``Pythagorean theorem:" 
\begin{equation}\label{sub10a}
C^2=
\left({}^{(1)}C^2+{}^{(2)}C^2+{}^{(3)}C^2\right)+\left({}^{(4)}C^2+
{}^{(5)}C^2\right)\,.
\end{equation}
\end{thm}

\subsection{Irreducible decompositions}
The decomposition (\ref{ORT3}) involves two scalars $S$ and $A$, two second order traceless tensors $P_{ij}$ and $Q_{ij}$, and a fourth order totally traceless tensor $R_{ijkl}$. 
Exactly the same types of tensors emerge in the harmonic decomposition that is widely used in elasticity theory. 
Such decomposition is generated by expressing the partial tensors in term of the harmonic polynomials, i.e.,  the polynomial solutions of the Laplace equation. The corresponding tensors are required to be completely symmetric and totally traceless. As it was demonstrated by  Backus \cite{Backus},  such a harmonic decomposition is not applicable in general in a space of the dimension greater than three. 

The most compact expression of this type was proposed by Cowin \cite{Cowin1989},  
\begin{eqnarray}\label{id1}
C^{ijkl}&=&ag^{ij}g^{kl}+b\left(g^{ik}g^{jl}+g^{il}g^{jk}\right)
+\left(g^{ij}{\hat A}^{kl}+g^{kl}{\hat A}^{ij}\right) +\nonumber\\
&&\left(g^{ik}{\hat B}^{jl}+g^{il}{\hat B}^{jk}+g^{jk}{\hat B}^{il}+g^{ik}{\hat B}^{jl}\right)+Z^{ijkl}\,.
\end{eqnarray}
 An alternative expression was proposed by Backus \cite{Backus}. It reads, see \cite{Baerheim},
\begin{eqnarray}\label{id2}
C^{ijkl}&=&H^{ijkl}+ \left(H^{ij}g^{kl}+H^{ik}g^{jl}+H^{il}g^{jk}+H^{jk}g^{il}
+H^{jl}g^{ik}+H^{kl}g^{ij}\right)+\nonumber\\
&&H\left(g^{ij}g^{kl}+g^{ik}g^{jl}+g^{il}g^{jk}\right)+\nonumber\\
&&
\left(h^{ij}g^{kl}+h^{kl}g^{ij}-\frac 12 h^{jl}g^{ik}-
\frac 12h^{ik}g^{jl}-\frac 12h^{jk}g^{il} -\frac 12h^{il}g^{jk} \right)
+\nonumber\\
&&
h\left(g^{ij}g^{kl}-\frac12g^{il}g^{jk}-\frac12 g^{ik}g^{jl}\right)
\end{eqnarray}
Let us compare these two expressions. First we observe that two totally traceless tensors must be equal to one  another, $H^{ijkl}=Z^{ijkl}$.
As for the scalar terms in Eq.(\ref{id1}), they are merely linear combinations of the corresponding terms in Eq.(\ref{id2}). Indeed, it is enough to take 
\begin{equation}\label{id4}
a=H+h \quad {\rm and}\quad b=H-(1/2) h\,. 
\end{equation}
Quite similarly, the traceless second order tensors of Eq.(\ref{id1}) are linear combinations of the corresponding terms of Eq.(\ref{id2}) with the identities 
\begin{equation}\label{id5}
{\hat A}^{kl}=H^{kl}+h^{kl} \quad {\rm and}\quad {\hat B}^{jl}=H^{jl}-(1/2)h^{jl}\,. 
\end{equation}
Both decompositions are irreducible under the action of the rotation group. The key difference between the two is that that the decomposition of Backus is also irreducible under the action of permutation group. Cowin's decomposition is reducible in this sense. 

Let us compare now the harmonic decomposition Eq.(\ref{id2}) to our decomposition as it is given in  Eq.(\ref{sub10}). We immediately identify 
\begin{eqnarray}\label{id6}
{}^{(1)}\!S^{ijkl}&=&H\left(g^{ij}g^{kl}+g^{ik}g^{jl}+
g^{il}g^{jk}\right), \label{id9} \\
{}^{(2)}\!S^{ijkl}&=& H^{ij}g^{kl}+H^{ik}g^{jl}+H^{il}g^{jk}+H^{jk}g^{il}
+H^{jl}g^{ik}+H^{kl}g^{ij}\,, \label{id8} \\
{}^{(3)}\!S^{ijkl}&=& H^{ijkl}\,, \label{id7}\\ 
{}^{(1)}\!A^{ijkl}&=& h\left(g^{ij}g^{kl}-\frac12g^{il}g^{jk}-\frac12 g^{ik}g^{jl}\right), \\
{}^{(2)}\!A^{ijkl}&=&h^{ij}g^{kl}+h^{kl}g^{ij}-\frac 12 h^{jl}g^{ik}-
\frac 12h^{ik}g^{jl}-\frac 12h^{jk}g^{il} -\frac 12h^{il}g^{jk} \,. \label{id10}
\end{eqnarray}
 We can straightforwardly derive the relations 
\begin{equation}\label{id11}
H=\frac 1{15} S\,,\qquad H^{ij}=\frac 17 P^{ij}\,, \qquad H^{ijkl}=R^{ijkl}\,,
\end{equation}
and
\begin{equation}\label{id12}
h=\frac 1{3} A\,,\qquad h^{ij}=-2 Q^{ij}\,.
\end{equation}

Thus the two decompositions are equivalent. The difference is that in Eq.(\ref{sub10}) the partial  tensors  are identified as elasticities themselves. These partial tensors generate the minimal direct sum  decomposition of the elasticity tensor space. Moreover. the corresponding subspaces are mutually orthogonal and  the squares of the tensors satisfy the ``Pythagorean theorem". We will see in the following section how those properties can be applied to the problem of quadratic invariants. 
\section{Linear and quadratic invariants}
\subsection{Linear invariants}
Each linear invariant of the elasticity tensor can be represented as 
\begin{equation}\label{lin1}
I=h_{ijkl}C^{ijkl}\,,
\end{equation}
where $h_{ijkl}$ is a tensor. 
Using the  decomposition (\ref{ORT3}), it can be rewritten as a sum of irreducible parts with different leading coefficients 
\begin{equation}\label{lin2}
I=\sum_{I=1}^5{}^{(I)}h_{ijkl}{}^{(I)}C^{ijkl}\,.
\end{equation}
We observe that the tensor $h_{ijkl}$ can be constructed only as a product of two components of the metric tensor, $h_{ijkl}\sim (g\cdot g)_{ijkl}$. Since the tensors $P_{ij}$, $Q_{ij}$, and  $R_{ijkl}$ are totally traceless they do not contribute to the sum in Eq.(\ref{lin2}). Consequently we are left with
\begin{equation}\label{lin2a}
I=\a S+\b A\,.
\end{equation} 
Hence, every linear invariant is represented as a linear combination of the  two basic linear invariants 
\begin{equation}\label{quad2}
L_1=S\stackrel{*}=\frac 13 \left(C_{iikk}+2 C_{ikik}\right)\,, \qquad L_2=A\stackrel{*}=\frac 13 \left(C_{iikk}- C_{ikik}\right)\,.
\end{equation}
Their independence can be seen from their explicit expressions. It follows also from the fact that $S$ and $A$ are related to two different subspaces ${}^{(1)}\S$ and ${}^{(1)}\A$. 
We readily obtain the expression of the linear invariants given in Eq.(\ref{intro2}),
\begin{equation}\label{quad3}
C_{ikik}=S-A\,, \qquad C_{iikk}=S+2A\,.
\end{equation}
\subsection{Quadratic invariants}
 Norris \cite{Norris} presented a generic quadratic invariant of the elasticity tensor in the form
 \begin{equation}\label{quad0}
f_{ijklpqrs}C^{ijkl}C^{pqrs}\,.
\end{equation}
Here $f_{ijklpqrs}$ is a numerical tensor. 
Using the irreducible decomposition (\ref{ORT3}) it can be  rewritten  as
\begin{equation}\label{quad1}
\sum_{I,J=1}^5{}^{(I,J)}\!f_{ijklpqrs}{}^{(I)}\!C^{ijkl}\,\,{}^{(J)}\!C^{pqrs}\,.
\end{equation}
The tensors with the  components ${}^{(I,J)}\!f_{ijklpqrs}$ can be constructed only as a product of four  components of the metric tensor, ${}^{(I,J)}\!f_{ijklpqrs}\sim (g\cdot g \cdot g \cdot g)_{ijklpqrs}$. 
Using the traceless property of the tensors $P_{ij}$, $Q_{ij}$, and $R_{ijkl}$, we can show that 
the quadratic invariants can be chosen uniquely as
\begin{equation}\label{quad3a}
Z_1=S^2\,, \qquad Z_2=AS\,, \qquad Z_3=A^2\,,
\end{equation}
\begin{equation}\label{quad4aa}
Z_4=P_{ij}P^{ij}\,, \qquad Z_5=P_{ij}Q^{ij}\,, \qquad Z_6=Q_{ij}Q^{ij}\,, \qquad {\rm and}\qquad Z_7=R_{ijkl}R^{ijkl}.
\end{equation}
It is clear that this set of invariants is complete. Indeed, when two tensors in (\ref{quad1}) are irreducibly decomposed in the form (\ref{ORT3}) with an arbitrary tensor $f_{ijklpqrs}$ constructed from the metric tensor and numbers, only the terms (\ref{quad3a}, \ref{quad4aa}) can appear. 
Moreover, these invariants are independent. It is due to the fact that they are taken from independent and even orthogonal subspaces of the elasticity tensor space. 
\subsection{Relations between two sets of quadratic invariants}
Let us display the quadratic invariants of the set (\ref{intro5})  in terms of $Z_I$. We present the details of calculations in the Appendix. 
For the quadratic invariants constructed from the linear ones, we have straightforwardly 
\begin{eqnarray}\label{rel1}
A_1^2&=&(S-A)^2=Z_1-2Z_2+Z_3\,,\\
A_1A_2&=&(S-A)(S+2A)=Z_1+Z_2-2Z_3\,,\\
A_2^2&=&(S+2A)^2=Z_1+4Z_2+4Z_3\,.
\end{eqnarray}
For the first invariant of Ting, we write
\begin{equation}\label{quad4}
B_1=C^{ijkl}C_{ijkl}=\sum_{I=1}^5{}^{(I)}\!C^{ijkl}\,\,\sum_{J=1}^5{}^{(J)}\!C_{ijkl}\,.
\end{equation}
Due to the orthogonality of the set, we are left here with 
\begin{equation}\label{quad4a}
B_1={}^{(1)}S^2+{}^{(2)}S^2+{}^{(3)}S^2+
{}^{(4)}A^2+{}^{(5)}A^2
\end{equation}
Let us list the expressions of these invariants
\begin{equation}\label{quad4ax}
{}^{(1)}S^2=\frac 15 S^2 \,, \qquad {}^{(2)}S^2=\frac 67 P^{ij}P_{ij} \,, \qquad {}^{(3)}S^2=R^{ijkl}R_{ijkl}\,,
\end{equation}
and
\begin{equation}\label{quad4bx}
{}^{(1)}A^2=A^2\,, \qquad {}^{(2)}A^2=12Q^{mn}Q_{mn}\,.
\end{equation}
Consequently,
\begin{equation}\label{quad4a6}
B_1=\frac 15 Z_1+Z_3+\frac 67Z_4+12Z_6+Z_7\,.
\end{equation}
The second invariant of Ting takes the form
\begin{equation}\label{quad6}
B_2=\frac 13Z_1 +\frac {4}3Z_2+\frac{4}{3}Z_3 +
Z_4-4Z_5+4Z_6\,.
\end{equation}
The first and the second invariants of   Ahmad read
\begin{equation}\label{quad6}
B_3=\frac 13Z_1 +\frac {1}3Z_2-\frac{2}{3}Z_3 +
Z_4-Z_5-2Z_6\,
\end{equation}
and 
\begin{equation}\label{quad7}
B_4=\frac 13Z_1 -\frac {2}3Z_2+\frac{1}{3}Z_3 +
Z_4+2Z_5+Z_6\,,
\end{equation}
respectively. From these expressions we see that the set of invariants (\ref{intro5}) is complete and the invariants are independent.  The same is true  for our set $Z_i$. 
We calculate also the additional invariant of Norris:
\begin{equation}\label{quad8}
B_5=C^{ijkl}C_{ikjl}=\frac 15 Z_1-\frac 1{2}Z_3+\frac 67 Z_4 -6Z_6+Z_7\,.
\end{equation}
Substituting the expressions (\ref{quad3a}--\ref{quad4aa}) we obtain the formula (\ref{intro8}) of Norris. 
\section{Applications: Invariants as characteristics of materials}
Since invariants of the elasticity tensor are independent of the coordinate system used in  specific measurements, they can be used as intrinsic characteristics of the materials. It is clear that linear independence is not enough for this goal. Indeed, although  the invariants $Z_1,Z_2$, and $Z_3$ are linear independent, they are related by a quadratic relation $Z_2^2=Z_1Z_3$. We will show that an intrinsic
meaning can be assigned to the five invariants that correspond to  different direct subspaces of the elasticity tensor space, namely 
 \begin{equation}\label{app1}
\{Z_1\,,Z_3\,,Z_4\,,Z_6\,,Z_7\}\,.
\end{equation}
All these invariants are positive.
\subsection{Cauchy relations and Cauchy factor}
In the early days of the elasticity theory,  Cauchy formulated a molecular model for elastic bodies, based on 15 independent elasticity constants. In this way 6 constraints, called {\it Cauchy relations}  were assumed. 
A lattice-theoretical analysis  shows, see \cite{Haus}, \cite{Leibfried} , that the Cauchy relations are valid provided the following conditions hold:
\begin{itemize}
 \item The interaction
forces between the  molecules of a crystal are central
forces; 
\item each  molecule is a center of
symmetry;
\item the interaction forces between the building blocks
of a crystal can be well approximated by a harmonic potential.
\end{itemize}
More recent discussions of the Cauchy relations can be found, e.g., in
[1], [3], [4], or [7]. Different  compact expressions of the Cauchy relations can be found in literature. For instance in  \cite{Haus},  they are presented as
\begin{equation}\label{cauchy-haus}
 C^{iijk}-C^{ijik}=0\,.
\end{equation}
An alternative form is widely  used, see \cite{Podio-Guidugli}, \cite{Weiner},  \cite{Cowin}, \cite{Campanella},
\begin{equation}\label{cauchy1a}
 C^{ijkl}-C^{ikjl}=0\,.
\end{equation}
The irreducible decomposition technique \cite{Cauchy}  yields 
\begin{equation}\label{cauchy2}
A^{ijkl}=0\,, \qquad {\rm or}\qquad Q_{ij}=0\,.
\end{equation}
As it was demonstrated experimentally already by Voigt,  the Cauchy relations do not hold even approximately.  Thus, elastic properties of the  generic anisotropic material is described by the whole set of   21 independent component. 
In fact, the situation  with the Cauchy relations is much more interesting, see \cite{Haus}. One can look for  the {\it deviation of the elasticity tensor from its Cauchy part.} As it was pointed out by Hauss\"uhl  \cite{Haus}, this deviation, even being a macroscopic characteristic,  can provide some important information about   microscopic structure of the material. To have such  deviation term we must have a unique proper decomposition of the elasticity tensor into two independent parts that can be referred to as Cauchy and non-Cauchy parts.

In \cite{Haus}, the deviation from the Cauchy part was presented by the value of the corresponding combination given in the left hand side of Eq.(\ref{cauchy-haus}). Such way of expression is valid only in the case when the non-Cauchy part is presented by only one component. Moreover this expression is dimension-full and depends on the choice of the coordinate system. 

With the use of the quadratic invariants we can introduce an invariant characteristic of deviation of a material from its Cauchy prototype. Due to the ``Pythagorean theorem" (\ref{ORT3}), we define the dimensionless quantity, which we will call the {\it Cauchy factor}
\begin{equation}\label{app4}
{\cal F}_{\rm Cauchy}=\sqrt{\frac {S^{ijkl}S_{ijkl}}{C^{ijkl}C_{ijkl}}}\,.
\end{equation}
Evidently, $0\le {\cal F}_{\rm Cauchy}\le 1$. A pure Cauchy material is determined by ${\cal F}_{\rm Cauchy}= 1$. For  ${\cal F}_{\rm Cauchy}= 0$, we have a hypothetic material without Cauchy part at all. Comparing two materials, we must conclude that a material with higher Cauchy factor has a microscopic structure closer to spherical symmetry. 

\subsection{Fedorov's problem}
In linear elasticity for anisotropic materials one must deal with a big set of elasticity constants.  But in some problems, the elastic body can only be  slightly different from an isotropic one. Fedorov \cite{Fedorov}, in a classical book on the propagation of elastic waves in anisotropic crystals,
has demonstrated how the anisotropic elastic tensor can be averaged over the 3-dimensional 
spatial directions in order to find some kind of isotropic approximation.

Recently Norris  \cite{Norris1}  took up
this program and defined an Euclidean distance function for solving the Fedorov problem in
a novel way. He succeeded in doing so and even extended the formalism for averaging the given set of elastic parameters relative to less symmetric classes.  The corresponding procedure can be outlined as follows:
\begin{itemize}
\item For a given 4th order elasticity tensor, one constructs the corresponding 2nd order Christoffel tensor, which is quadratic in the wave vector.
\item One consider the $C^2$ norm of the difference between the given anisotropic Christoffel tensor and a generic isotropic one. The isotropic Christoffel tensor is taken with two unknown parameters.
\item Since Christoffel tensor is quadratic in the wave vector ${\bf n}$,   the mentioned  norm is quartic ${\bf n}$ likewise. 
Its  average  is computed in space directions.
\item The resulting expression is left to be a function of two isotropic parameters. Its minimization is applied and the resulting pair of isotropic parameters is derived. 
\end{itemize}
 The result of this consideration  is given in  \cite{Norris1} as
\begin{equation}\label{fed}
\kappa\stackrel{*}=\frac 19 \,C_{iijj},\qquad \mu\stackrel{*}=\frac 1{10}\,C_{ijij}-\frac 1{30}\,C_{iijj}\,.
\end{equation}

In \cite{Norris},  \cite{Norris1} and \cite{Moakher-Norris}, Norris explained that  minimizing a Euclidean distance function is equivalent to projecting the tensor of elastic stiffness onto the appropriate symmetry. 

Let us consider an elasticity tensor of 21 independent components. It is irreducibly decomposed to the sum of five independent pieces. Two scalar pieces, namely ${}^{(1)}\!S^{ijkl}$ and ${}^{(1)}\!A^{ijkl}$ has a special property: They are invariant under arbitrary $SO(3)$ transformation. 
Using the direct sum of the corresponding subspaces, we can construct a subspace
\begin{equation}\label{iso1}
{\cal ISO}={}^{(1)}{\cal S}\otimes {}^{(1)}{\cal A}\,.
\end{equation}
This 2-dimensional subspace  is $SO(3)$ invariant and orthogonal to all other subspaces of an  arbitrary elasticity tensor. Evidently it must be identified as an {\it isotropic part} of an elasticity tensor. Consequently we constructed an isotropic part of the generic elasticity tensor in the form
\begin{equation}\label{iso1}
{}^{\rm(iso)}C^{ijkl}={}^{(1)}\!S^{ijkl}+{}^{(1)}\!A^{ijkl}\,,
\end{equation}
or, explicitly,
\begin{eqnarray}\label{iso2}
^{\rm(iso)}C^{ijkl}=\frac 1{15}S \left(g^{ij}g^{kl}+g^{ik}g^{jl}+g^{il}g^{jk}\right)+\frac 16
A\left(2g^{ij}g^{kl}-g^{il}g^{jk}-g^{ik}g^{jl}\right) \,.
\end{eqnarray}
Let us compare this expression to the standard representation of an isotropic material in terms of the Lam\'e moduli $\lambda$ and $\mu$
\begin{equation}\label{iso3}
^{\rm(iso)}C^{ijkl}=\lambda g^{ij}g^{kl}+\mu\left(g^{ik}g^{jl}+g^{il}g^{jk}\right)\,.
\end{equation}
We derive the {\it effective Lame moduli} for an anisotropic material
\begin{equation}\label{iso4}
\lambda=\frac 1{15}S+\frac 13 A\,,\qquad \mu=\frac 1{15}S-\frac 16A\,.
\end{equation}
Recall that 
\begin{equation}\label{iso5}
S\stackrel{*}=\frac 13\left(C_{iikk}+2C_{ikik}\right)\,,\qquad 
A\stackrel{*}=
\frac 13\left(C_{iikk}-C_{ikik}\right)\,.
\end{equation}
Substituting these expressions into equations (\ref{iso4}), we derive
\begin{equation}\label{iso8}
\lambda\stackrel{*}=\frac 2{15}\,C_{iijj}-\frac 1{15}\,C_{ijij}\,,\qquad \mu\stackrel{*}=\frac 1{10}\,C_{ijij}-\frac 1{30}\,C_{iijj}\,.
\end{equation}
With the bulk constant $\kappa=\lambda+(2/3)\mu$ we recover both    expressions given in Eq.(\ref{fed}). 

In order to express  the deviation of the given anisotropic material from its effective isotropic prototype, one uses the distance between two tensors. This quantity is dimensionful and depends on the average magnitude of the elasticity tensor. Instead, we define the {\it isotropy factor} of an anisotropic material in the form
\begin{equation}\label{iso9}
{\cal F}_{\rm iso}=\sqrt{\frac {^{\rm(iso)}C^2}{C_{ijkl}C^{ijkl}}}=\sqrt{\frac {{}^{(1)}S_{ijkl}{}^{(1)}S^{ijkl}+{}^{(1)}A_{ijkl}{}^{(1)}A^{ijkl}}{C_{ijkl}C^{ijkl}}}\,.
\end{equation}
In terms of the constants $S$ and $A$ it reads
\begin{equation}\label{iso10}
{\cal F}_{\rm iso}=\sqrt{\frac {S^2+5A^2}{5C_{ijkl}C^{ijkl}}}=\sqrt{\frac {Z_1+5Z_3}{ Z_1+5Z_3+ ({30}/7)Z_4+60Z_6+5Z_7}}\,.
\end{equation}
We observe that $0\le{\cal F}_{\rm iso}\le 1$. It is equal to one for pure isotropic materials and equal to zero for some hypothetic material without isotropic part, i.e., in the case when the effective Lam\'e moduli vanish. 

\subsection{Irreducibility factors}
As a natural extension of the Cauchy and the isotropy factors   described above, we introduce  dimensionless numerical factors that describe the contribution of the irreducible pieces to the  elasticity tensor. For the 5 irreducible parts ${}^{(I)}C^{ijkl}$ with $I=1,\cdots,5$, we define the {\it irreducibility factors}
 \begin{equation}\label{iso10a}
{}^{(I)}{\cal F}_{\rm irr}=\sqrt{\frac {{}^{(I)}C_{ijkl}{}^{(I)}C^{ijkl}}{C_{ijkl}C^{ijkl}}}\,.
\end{equation}
In particular, the Cauchy factor is expressed as
\begin{equation}\label{iso10b}
{\cal F}_{\rm Cauchy}=\sqrt{\frac{{}^{(1)}A_{ijkl}{}^{(1)}A^{ijkl}+{}^{(2)}A_{ijkl}{}^{(2)}A^{ijkl}}{C_{ijkl}C^{ijkl}}}=\sqrt{{}^{(4)}{\cal F}^2_{\rm irr}+{}^{(5)}{\cal F}^2_{\rm irr}}\, 
\end{equation}
and the isotropy factor  as 
\begin{equation}\label{iso10C}
{\cal F}_{\rm iso}=\sqrt{\frac{{}^{(1)}S_{ijkl}{}^{(1)}S^{ijkl}+{}^{(1)}A_{ijkl}{}^{(1)}A^{ijkl}}{C_{ijkl}C^{ijkl}}}=\sqrt{{}^{(1)}{\cal F}^2_{\rm irr}+{}^{(4)}{\cal F}^2_{\rm irr}}\,.
\end{equation}
\section{Cubic crystals}
\subsection{Definition}
Cubic crystals are described by three independent elasticity
constants. In a properly chosen coordinate system, they can be put,
see Nayfeh \cite{Nayfeh}, into the following Voigt matrix:
\begin{equation}\label{cub-voigt}
\begin{bmatrix}
C^{1111} & C^{1122} & C^{1133} & C^{1123} & C^{1131} & C^{1112} \\
* & C^{2222} & C^{2233} & C^{2223} & C^{2231} & C^{2212} \\
* & * & C^{3333} & C^{3323} & C^{3331} & C^{3312} \\
* & * & * & C^{2323} & C^{2331} & C^{2312} \\
* & * & * & * & C^{3131} & C^{3112} \\
* & * & * & * & * & C^{1212}
     \end{bmatrix} \equiv \begin{bmatrix}
  C^{11} & C^{12} & C^{12} & 0 & 0 & 0 \\
* & C^{11} & C^{12} & 0 & 0 & 0 \\
* & * & C^{11} & 0 & 0 & 0 \\
* & * & * & C^{66} & 0 & 0 \\
* & * & * & * & C^{66} & 0 \\
* & * & * & * & * & C^{66} \end{bmatrix}\,.
\end{equation}
Taking into account the multiplicities of the elements exhibited in  (\ref{cub-voigt}), we calculate
 \begin{equation}\label{cub-C2}
C^2=C_{ijkl}C^{ijkl}=3\left(C^{11}\right)^2+6\left(C^{12}\right)^2+12\left(C^{66}\right)^2\,.
\end{equation}
\subsection{$S_4$-decomposition}
We decompose (\ref{cub-voigt}) irreducibly and find the Cauchy part
\begin{equation}\label{cub-sym}
S^{ijkl}=\begin{bmatrix}
S^{1111} & S^{1122} & S^{1133} & S^{1123} & S^{1131} & S^{1112} \\
* & S^{2222} & S^{2233} & S^{2223} & S^{2231} & S^{2212} \\
* & * & S^{3333} & S^{3323} & S^{3331} & S^{3312} \\
* & * & * & S^{2323} & S^{2331} & S^{2312} \\
* & * & * & * & S^{3131} &S^{3112} \\
* & * & * & * & * & S^{1212}
     \end{bmatrix} = \begin{bmatrix}
{\a} & {\b} & {\b} & 0 & 0 & 0 \\
* & {\a} & {\b} &  0 & 0 & 0 \\
* & * & {\a} & 0 & 0 & 0 \\
* & * & * & {\b} & 0 & 0 \\
* & * & * & * & {\b} & 0 \\
* & * & * & * & * & {\b}
     \end{bmatrix} \,,\end{equation}\
where
\begin{equation}\label{cub-sym1}
{\a}=C^{11},\quad {\b}=\frac 13 \left(C^{12}+2C^{66}\right)\,,
\end{equation}
The square of this tensor takes the value
 \begin{equation}\label{cub-S2}
\widetilde{S}^2=S_{ijkl}S^{ijkl}=3\a^2+18\b^2=3\left(C^{11}\right)^2+2\left(C^{12}+2C^{66}\right)^2\,.
\end{equation}
The non-Cauchy part is represented by
\begin{equation}\label{cub-ant}
  A^{ijkl}=\begin{bmatrix}
A^{1111} & A^{1122} & A^{1133} & A^{1123} & A^{1131} & A^{1112} \\
* & A^{2222} & A^{2233} & A^{2223} & A^{2231} & A^{2212} \\
* & * & A^{3333} & A^{3323} & A^{3331} & A^{3312} \\
* & * & * & A^{2323} & A^{2331} & A^{2312} \\
* & * & * & * & A^{3131} &A^{3112} \\
* & * & * & * & * & A^{1212}
     \end{bmatrix} =  \begin{bmatrix}
    0 & 2{\g} & 2{\g} & 0 & 0 & 0 \\
    * & 0 & 2{\g} & 0 & 0 & 0 \\
    * & * & 0 & 0 & 0 & 0 \\
    * & * & * & -{\g} & 0 & 0 \\
    * & * & * & * & -{\g} & 0 \\
    * & * & * & * & * &-{\g} \end{bmatrix} \,,\end{equation}\
where
\begin{equation}\label{cub-sym1}
 {\g}=\frac 13 \left(C^{12}-C^{66}\right)\,. 
\end{equation}
Its square reads
 \begin{equation}\label{cub-A2}
\widetilde{A}^2=A_{ijkl}A^{ijkl}=36\g^2 =4\left(C^{12}-C^{66}\right)^2\,.
\end{equation}
We can straightforwardly check the identities 
 \begin{equation}\label{cub-A2x}
\widetilde {A}\cdot \widetilde{S}=0\,,\end{equation}
and 
\begin{equation}\label{cub-A2xx}
C^2=\widetilde{S}^2+\widetilde{A}^2=3\a^2+18\b^2+36\g^2\,.
\end{equation}
The Cauchy factor is expressed as 
 \begin{equation}\label{cub-F-Cauchy}
{\cal F}_{\rm Cauchy}=\sqrt{\frac{\a^2+6\b^2}{\a^2+6\b^2+12\g^2}}\,,
\end{equation}
or
 \begin{equation}\label{cub-F-Cauchy-1}
{\cal F}_{\rm Cauchy}= \sqrt{\frac{3\left(C^{11}\right)^2+2\left(C^{12}+2C^{66}\right)^2}{3\left(C^{11}\right)^2+6\left(C^{12}\right)^2+12\left(C^{66}\right)^2}}\,.
\end{equation}
For the Cauchy relation we have
\begin{equation}\label{cub-Cauchy}
{\g}=0\,,\qquad{\rm{or}}\qquad C^{12}=C^{66}\,.
\end{equation}
In this case, Eq.(\ref{cub-F-Cauchy}) yields $ {\cal F}_{\rm Cauchy}=1$. 

For cubic crystals, the Cauchy factor in Eqs.(\ref{cub-F-Cauchy},\ref{cub-F-Cauchy-1}) depends on two independent parameters and thus allows 3-dimensional visualization, see Fig. 1 and Fig 2.
\begin{figure}[h!]
{
\includegraphics[width=7.5cm]{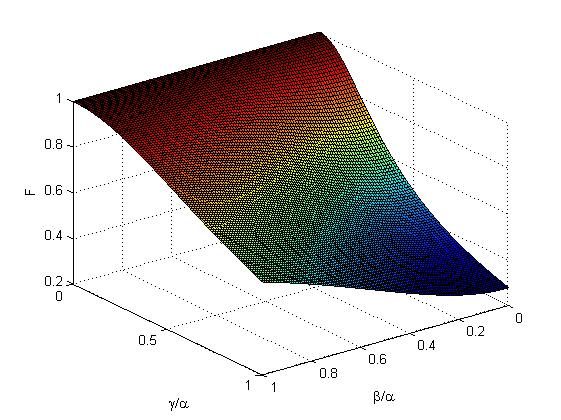}
\includegraphics[width=7.5cm]{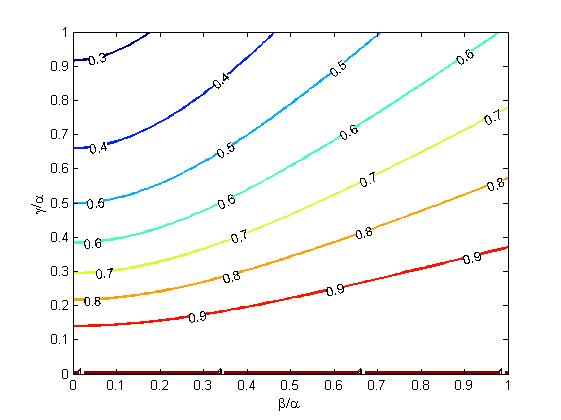}}
\caption[]{Functional dependence of the Cauchy factor $ {\cal F}_{\rm Cauchy}$ with respect to the parameters ${\b}/{\a}$ and ${\g}/{\a}.$ The pure Cauchy materials are depicted by the straight line lying on the axis $\g=0$. }
\end{figure}
\begin{figure}[h!]
{
\includegraphics[width=7.5cm]{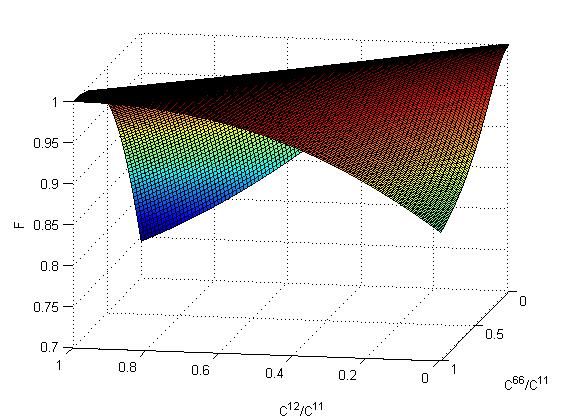}
\includegraphics[width=7.5cm]{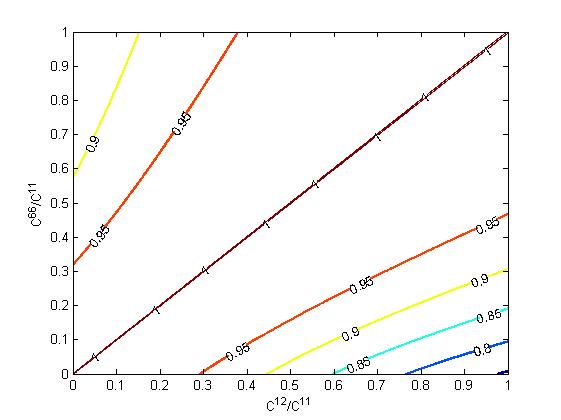}}
\caption[]{Functional dependence of the Cauchy factor $ {\cal F}_{\rm Cauchy}$ with respect to the parameters ${C^{12}}/{C^{11}}$ and ${C^{66}}/{C^{11}}.$ The pure Cauchy materials are depicted by the straight line  $C^{12}=C^{66}$.}
\end{figure}
\subsection{$SO(3)$-decomposition}
Let us determine  the $SO(3)$ components of the totally symmetric tensor. Using Eq.(\ref{S2}) we derive from (\ref{cub-sym}) 
  \begin{equation}\label{cub-S}
S^{ij}=(\a+2\b)g^{ij} =
\left(C^{11}+\frac 23C^{12}+\frac 43 C^{66}\right) g^{ij}\,.\end{equation}
Consequently,
 \begin{equation}\label{cub-S2}
S=3(\a+2\b)=3C^{11}+2C^{12}+4 C^{66}\,,\end{equation}
and
\begin{equation}\label{cub-P}
P^{ij}=0\,.\end{equation}
The total traceless remainder can be expressed  as 
\begin{equation}\label{cub-R}
R^{ijkl}=\begin{bmatrix}
R^{1111} & R^{1122} & R^{1133} & R^{1123} & R^{1131} & R^{1112} \\
* & R^{2222} & R^{2233} & R^{2223} & R^{2231} & R^{2212} \\
* & * & R^{3333} & R^{3323} & R^{3331} & R^{3312} \\
* & * & * & R^{2323} & R^{2331} & R^{2312} \\
* & * & * & * & R^{3131} &R^{3112} \\
* & * & * & * & * & R^{1212}
     \end{bmatrix} = \frac{\a-3\b}5\begin{bmatrix}
2 & -1 & -1 & 0 & 0 & 0 \\
* & 2 & -1 &  0 & 0 & 0 \\
* & * & 2 & 0 & 0 & 0 \\
* & * & * & -1 & 0 & 0 \\
* & * & * & * & -1 & 0 \\
* & * & * & * & * & -1
     \end{bmatrix} \,,\end{equation}\
The two non-zero quadratic invariants read
\begin{equation}\label{cub-S-qua}
S^2=9(\a+2\b)^2=\left(3C^{11}+2C^{12}+4C^{66}\right)^2\,\end{equation}
and
\begin{equation}\label{cub-R-qua}
R^2=\frac {6}{5}(\a-3\b)^2=\frac  {6}{5}\left(C^{11}-C^{12}-C^{66}\right)^2\,.\end{equation}
We check the formula
\begin{equation}\label{cub-S-check}
{\tilde S}^2=\frac {1}{5}S^2+R^2\,,\end{equation}
that must hold with the correspondence with the expressions (\ref{quad4ax},\ref{quad4bx}) of the first Ting's invariant. 

Let us turn now to the asymmetric part $A^{ijkl}$.  Eq.(\ref{cub-ant}) yields 
\begin{equation}\label{cub-delta}
\Delta^{ij}=2\g g^{ij}\,.\end{equation}
 Consequently,
\begin{equation}\label{cub-A1}
A=6\g\,,\qquad Q^{ij}=0\,.\end{equation}
Thus,
\begin{equation}\label{cub-A1}
\widetilde{A}^2=A^2=36\g^2=4\left(C^{12}-C^{66}\right)^2\,.\end{equation}
Accordingly, the isotropic factor takes the form
\begin{equation}\label{iso10a}
{\cal F}_{\rm iso}=\sqrt{\frac {(1/5)S^2+A^2}{C_{ijkl}C^{ijkl}}}=
\sqrt{\frac {(3/5)(\a+2\b)^2+12\g^2}{\a^2+6\b^2+12\g^2}}\,,
\end{equation}
or 
\begin{equation}\label{iso11b}
{\cal F}_{\rm iso}
=\sqrt{\frac {\left(3C^{11}+2C^{12}+4C^{66}\right)^2+20\left(C^{12}-C^{66}\right)^2}{15\left(\left(C^{11}\right)^2+2\left(C^{12}\right)^2+
4\left(C^{66}\right)^2\right)}}\,.
\end{equation}
It is well known that an isotropic medium can be described as a cubic crystal with $C^{66}=(1/2)(C^{11}-C^{12})$. It is equivalent to $\a=3\b$. When this relation is substituted into (\ref{iso10a}), we obtain ${\cal F}_{\rm iso}=1$. 
In Fig. 3 and Fig 4. we present the functional dependence of the isotropy factor for a cubic crystal with respect to the homogeneous fractions of its parameters. 
\begin{figure}[h!]
{
\includegraphics[width=7.5cm]{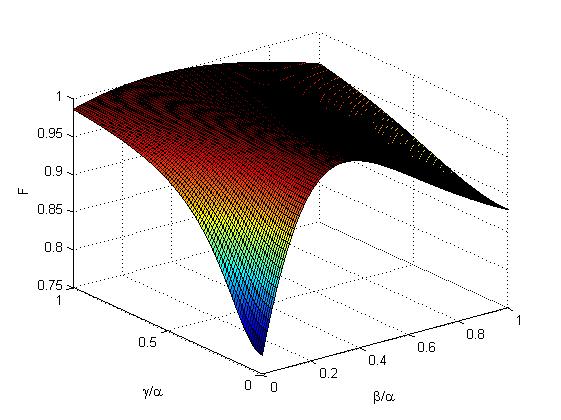}
\includegraphics[width=7.5cm]{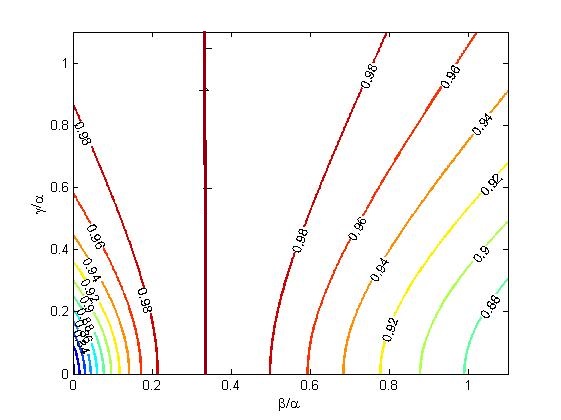}}
\caption[]{Functional dependence of the isotropy factor $ {\cal F}_{\rm iso}$  is depicted with respect to the variables ${\b}/{\a}$ and ${\g}/{\a}.$ The straight line with ${\b}/{\a}=1/3$ corresponds to the pure isotropic materials.}
\end{figure}
\begin{figure}[h!]
{
\includegraphics[width=7.5cm]{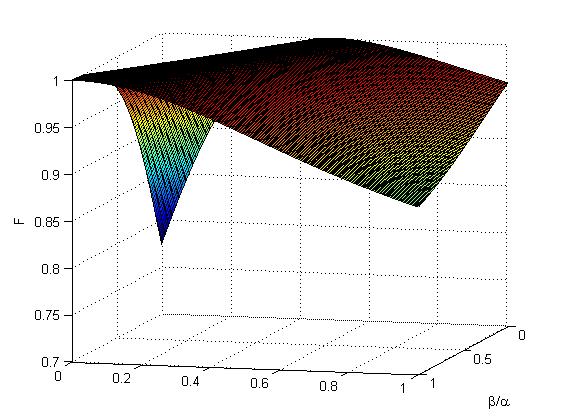}
\includegraphics[width=7.5cm]{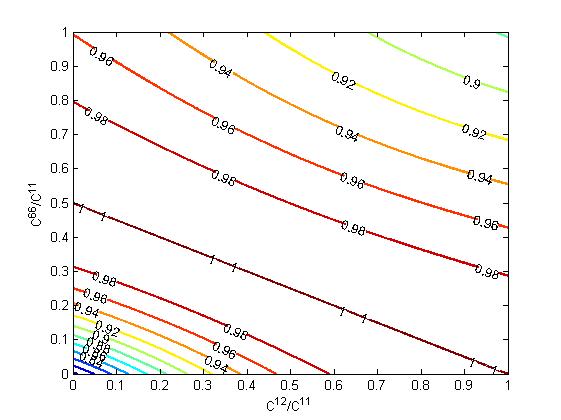}}
\caption[]{Functional dependence of the isotropy factor $ {\cal F}_{\rm iso}$  with respect to the variables ${C^{12}}/{C^{11}}$ and ${C^{66}}/{C^{11}}$. The straight line $C^{66}=(1/2)(C^{11}-C^{12})$ correspond to the isotropic medium.}
\end{figure}
\subsection{Cubic averaging}
We consider now a problem of averaging of an arbitrary elastic material by a cubic symmetry prototype. Let in some coordinate system a material be described by a full set of 21 elastic parameters $C^{ijkl}$. We are looking for a cubic crystal that is mostly close to our material with respect to the Euclidean metric. We assume that in the same coordinate system the cubic crystal is represented in the canonical form (\ref{cub-voigt}). Since for cubic crystal $P^{ij}=Q^{ij}=0$, its elasticity tensor is left  
\begin{equation}\label{cub1}
C^{ijkl}_{\rm cub}= {}^{(1)}S^{ijkl}_{\rm cub}+{}^{(3)}S^{ijkl}_{\rm cub}+{}^{(1)}A^{ijkl}_{\rm cub},
\end{equation}
where
\begin{eqnarray}\label{cub2}
{}^{(1)}S^{ijkl}_{\rm cub}&=&\frac 1{15} m\left(g^{ij}g^{kl}+g^{ik}g^{jl}+g^{il}g^{jk}\right) ,\\
{}^{(1)}\!A^{ijkl}_{\rm cub}&=&\frac 16
n\left(2g^{ij}g^{kl}-g^{il}g^{jk}-g^{ik}g^{jl}\right)\,,\\
{}^{(3)}S^{ijkl}_{\rm cub}&=&k\sigma^{ijkl}\,.
\end{eqnarray}
Here $\sigma^{ijkl}$ is a set of parameters (not a tensor) that is represented in the $6\times 6$ notations by the constant matrix in  the right hand side of Eq.(\ref{cub-R}). 
Thus $m,k,n$ are our unknown variables. 
The square distance between two elasticity tensors is given by
\begin{equation}\label{cub3}
\left(C^{ijkl}-C^{ijkl}_{\rm cub}\right)^2= K_1(S-m)^2+K_2(A-n)^2+(R^{ijkl}-k\sigma^{ijkl})^2+K_3\,,
\end{equation}
where $K_1,K_2,K_3$ are positive numerical constants. 
This expression reaches its minimal value for 
 \begin{equation}\label{cub3}
m=S\,, \qquad n=A\,,
\end{equation}
and 
 \begin{equation}\label{cub4}
(R^{ijkl}-k\sigma^{ijkl})\sigma_{ijkl}=0\,,
\end{equation}
We have 
 \begin{equation}\label{cub5}
k=\frac {R^{ijkl}\sigma_{ijkl}}{\sigma^{ijkl}\sigma_{ijkl}}\,,
\end{equation}
Using the expression of $\sigma^{ijkl}$ listed in Eq.(\ref{cub-R}),
we derive
 \begin{equation}\label{cub6}
k=\frac {R^{1111}+R^{2222}+R^{3333}}6\,.
\end{equation}
Consequently the parameters of the cubic crystal prototype take the values
\begin{eqnarray}
\a&=&\frac 15 S+2k=\frac 13\left(C^{1111}+C^{2222}+C^{3333}\right) -\frac 2{15}S\\
\b&=&\frac 1{15}S-k=-\frac 16\left(C^{1111}+C^{2222}+C^{3333}\right)  -\frac 7{30}S  \\
\g&=&\frac 16 A\,.
\end{eqnarray}
\section{Conclusion}
In the framework of the irreducible decomposition of elasticity tensor, we studied the problem of its quadratic invariants. Since this decomposition is orthogonal, the invariants emerge in a natural and systematic way. Their independence and completeness follow straightforwardly from the direct sum decomposition of the tensor space. For arbitrary anisotropic materials, we defined the Cauchy  factor as  a dimensionless measure of a closeness to a pure Cauchy material. Quite similarly, we defined isotropy factor as a measure for a closeness to an isotropic prototype of a given material. The irreducible factors are defined in order to characterize the  contributions of different irreducible parts of an anisotropic elasticity tensor. This formalism can be useful for various elasticity problems:
\begin{itemize}
\item Elasticity wave propagation \cite{Alshits};
\item complete set of anisotropy invariants, see \cite{Xiao};
\item material symmetries and wavefront symmetries \cite{Bona1}, \cite{Bona2};
\item averaging of anisotropic material by a higher symmetric prototype \cite{Moakher-Norris}, \cite{Norris1}.  
\end{itemize}

\section*{Acknowledgments}
I would like to thank F.-W.Hehl (Cologne/Columbia, MO) and A. Norris (Rutgers) for most 
helpful discussion and comments.

\appendix
\section{Calculating the relations between invariants}
\subsection{The first Ting invariant}
 For the first invariant of Ting, we write
\begin{equation}\label{quad4}
B_1=C^{ijkl}C_{ijkl}=\sum_{I=1}^5{}^{(I)}\!C^{ijkl}\,\,\sum_{J=1}^5{}^{(J)}\!C_{ijkl}\,.
\end{equation}
Due to the orthogonality of the decomposition,  we are left  with 
\begin{equation}\label{quad4a}
B_1={}^{(1)}S^2+{}^{(2)}S^2+{}^{(3)}S^2+
{}^{(1)}A^2+{}^{(2)}A^2
\end{equation}
We calculate step by step
\begin{eqnarray}\label{quad4a1}
{}^{(1)}S^2&=&\frac 1{225}S^2\left(g^{ij}g^{kl}+g^{ik}g^{jl}+g^{il}g^{jk}\right)^2=\frac 15 S^2=\frac 15Z_1\,,\\
\label{quad4a2}
{}^{(2)}S^2&=&\frac 1{49}\left(P^{ij}g^{kl}+P^{ik}g^{jl}+P^{il}g^{jk}+P^{jk}g^{il}
+P^{jl}g^{ik}+P^{kl}g^{ij}\right)^2=\frac 67 P^{ij}P_{ij}=\frac 67Z_4\,,\\
\label{quad4a3}
{}^{(3)}S^2&=&R^{ijkl}R_{ijkl}=Z_7\,,\\
\end{eqnarray}
and
\begin{eqnarray}\label{quad4a4}
{}^{(1)}A^2&=&\frac 1{36}A^2\left(2g^{ij}g^{kl}-g^{il}g^{jk}-g^{ik}g^{jl}\right)^2=A^2=Z_3\\
{}^{(2)}A^2&=&\frac 14\left(\epsilon^{ikm}\epsilon^{jln}+
\epsilon^{ilm}\epsilon^{jkn}\right)\Big(\epsilon_{ikp}\epsilon_{jlq}+
\epsilon_{ilp}\epsilon_{jkq}\Big)Q_{mn}Q^{pq}=12Q^{mn}Q_{mn}
=12Z_6\,.
\label{quad4a5}
\end{eqnarray}
Consequently, the first invariant of Ting reads
\begin{equation}\label{quad4a6}
B_1=\frac 15 Z_1+Z_3+\frac 67Z_4+12Z_6+Z_7\,.
\end{equation}
\subsection{The second Ting invariant}
 For the second invariant of Ting, $B_2$, we first calculate the trace
\begin{equation}\label{quad5}
C^i{}_{ikl}=g^{ij}C_{ijkl}=g^{ij}\left(S_{ijkl}+
{}^{(1)}\!A_{ijkl}+{}^{(2)}\!A_{ijkl}\right)\,.
\end{equation}
Here,
\begin{equation}\label{quad6}
g^{ij}S_{ijkl}=S_{kl}=P_{kl}+\frac 13 Sg_{kl}\,,
\end{equation}
\begin{equation}\label{quad7}
g^{ij}{}^{(1)}\!A_{ijkl}=g^{ij}\left(g_{ik}Q_{jl}+g_{jk}Q_{il}+g_{il}Q_{jk}+
g_{jl}Q_{ik}-2g_{kl}Q_{ij}-2g_{ij}Q_{kl}\right)= -2Q_{kl}\,,
\end{equation}
and
\begin{equation}\label{quad8}
g^{ij}\,{}^{(2)}\!A_{ijkl}=\frac 16
Ag^{ij}\left(2g_{ij}g_{kl}-g_{il}g_{jk}-g_{ik}g_{jl}\right)=\frac 23Ag_{kl}\,.
\end{equation}
Hence, 
\begin{equation}\label{quad9}
C^i{}_{ikl}=P_{kl}-2Q_{kl} +\frac 13\left( S+2A\right)g_{kl}\,.
\end{equation}
Consequently, the second invariant of Ting reads
\begin{equation}\label{quad10}
B_2=C^i{}_{ikl}C_j{}^{jkl}=P_{kl}P^{kl}-4P_{kl}Q^{kl}+4Q_{kl}Q^{kl}+\frac 13\left(S+2A\right)^2\,,
\end{equation}
or
\begin{equation}\label{quad11}
B_2=\frac 13Z_1 +\frac {4}3Z_2+\frac{4}{3}Z_3 +
Z_4-4Z_5+4Z_6\,.
\end{equation}
\subsection{The first Ahmad invariant} 
For the first invariant of  Ahmad, $B_3$, we need the trace
\begin{equation}\label{quad12}
C^j{}_{kjl}=g^{ij}C_{ikjl}=g^{ij}\left(S_{ikjl}+
{}^{(1)}\!A_{ikjl}+{}^{(2)}\!A_{ikjl}\right)\,.
\end{equation}
Here,
\begin{equation}\label{quad13}
g^{ij}S_{ikjl}=S_{kl}=P_{kl}+\frac 13 Sg_{kl}\,,
\end{equation}
\begin{equation}\label{quad14}
g^{ij}\,{}^{(1)}\!A_{ikjl}=\frac 16
Ag^{ij}\left(2g_{ik}g_{jl}-g_{il}g_{jk}-g_{ij}g_{kl}\right)=-\frac 13Ag_{kl}\,.
\end{equation}
and
\begin{equation}\label{quad15}
g^{ij}\,{}^{(2)}\!A_{ikjl}=g^{ij}\left(g_{ij}Q_{kl}+g_{jk}Q_{il}+g_{il}Q_{jk}+
g_{kl}Q_{ij}-2g_{jl}Q_{ik}-2g_{ik}Q_{jl}\right)= Q_{kl}\,,
\end{equation}
Consequently,
\begin{equation}\label{quad16}
C^j{}_{kjl}=P_{kl}+Q_{kl}+\frac 13 g_{kl}(S-A)\,.
\end{equation}
Hence using (\ref{quad9}) and (\ref{quad16}) we get
\begin{eqnarray}\label{quad17}
B_3&=&\left(P_{kl}+Q_{kl}+\frac 13 g_{kl}(S-A)\right)\left(P^{kl}-2Q^{kl} +\frac 13\left( S+2A\right)g^{kl}\right)\nonumber\\
&=&P_{kl}P^{kl}-P_{kl}Q^{kl}-2Q_{kl}Q^{kl}+\frac 13(S+2A)(S-A)\,,
\end{eqnarray}
or
\begin{equation}\label{quad18}
B_3=\frac 13Z_1 +\frac {1}3Z_2-\frac{2}{3}Z_3 +
Z_4-Z_5-2Z_6\,.
\end{equation}
\subsection{The second Ahmad invariant} 
This invariant is obtained by the use of the formula (\ref{quad16}),
\begin{eqnarray}\label{quad19}
B_3&=&\left(P^{kl}+Q^{kl}+\frac 13 g^{kl}(S-A)\right)\left(P_{kl}+Q_{kl}+\frac 13 g_{kl}(S-A)\right)\nonumber\\
&=&P_{kl}P^{kl}+2P_{kl}Q^{kl}+Q_{kl}Q^{kl}+\frac 13(S-A)^2\,,
\end{eqnarray}
or
\begin{equation}\label{quad20}
B_4=\frac 13Z_1 -\frac {2}3Z_2+\frac{1}{3}Z_3 +
Z_4+2Z_5+Z_6\,.
\end{equation}
\subsection{The Norris invariant} 
We put the invariant of Norris  first  in the form
\begin{equation}\label{quad21}
B_5=C^{ijkl}C_{ikjl}=\left(S^{ijkl}+A^{ijkl}\right)\left(S_{ikjl}+A_{ikjl}\right)=
S^{ijkl}S_{ijkl}+A^{ijkl}A_{ikjl}\,.
\end{equation}
Using (\ref{quad4a1},\ref{quad4a2},\ref{quad4a3}), we have
\begin{equation}\label{quad22}
S^{ijkl}S_{ikjl}=S^{ijkl}S_{ijkl}=\frac 15 Z_1+\frac 67 Z_4+Z_7\,.
\end{equation}
We observe
\begin{equation}\label{quad23}
A^{ijkl}A_{ikjl}=\left({}^{(1)}\!A^{ijkl}+{}^{(2)}\!A^{ijkl}\right)
\left({}^{(1)}\!A_{ikjl}+{}^{(2)}\!A_{ikjl}\right)={}^{(1)}\!A_{ijkl}\,{}^{(1)}\!A^{ikjl}+{}^{(2)}\!A_{ijkl}\,{}^{(2)}\!A^{ikjl}\,.
\end{equation}
Thus we find
\begin{eqnarray}\label{quad24}
{}^{(1)}\!A_{ijkl}\,{}^{(1)}\!A^{ikjl}=\frac 1{36}A^2\big(2g_{ij}g_{kl}-g_{il}g_{jk}-g_{ik}g_{jl}\big)\big(2g^{ik}g^{jl}-g^{il}g^{jk}-g^{ij}g^{kl}\big)=-\frac 1{2}A^2\,,
\end{eqnarray}
\begin{eqnarray}\label{quad25}
{}^{(2)}\!A_{ijkl}\,{}^{(2)}\!A^{ikjl}&=&\big(g_{ik}Q_{jl}+g_{jk}Q_{il}+g_{il}Q_{jk}+
g_{jl}Q_{ik}-2g_{kl}Q_{ij}-2g_{ij}Q_{kl}\big)\times\nonumber\\
&&\big(g^{ij}Q^{kl}+g^{jk}Q^{il}+g^{il}Q^{jk}+
g^{kl}Q^{ij}-2g^{jl}Q^{ik}-2g^{ik}Q^{jl}\big)\nonumber\\
&=&-6Q_{ij}Q^{ij}\,.
\end{eqnarray}
Consequently,
\begin{equation}\label{quad26}
B_5=C^{ijkl}C_{ikjl}=\frac 15 Z_1-\frac 1{2}Z_3+\frac 67 Z_4 -6Z_6+Z_7\,.
\end{equation}


\begin{thebibliography}{99}

\bibitem{Ahmad} F. Ahmad (2002) Invariants and structural invariants of the anisotropic elasticity tensor, {\it Quarterly J. Mech.
Appl. Math.}, {\bf 55}(4), 597--606.

\bibitem{Alshits} Alshits, V. I. and Lothe, J. (2004) Some basic properties of bulk elastic waves in anisotropic media, {\it Wave Motion} {\bf 40}, 297--313.

\bibitem{Backus} G. Backus (1970) A geometrical picture of anisotropic elastic tensors, {\it Rev. Geophys. Space Phys.} {\bf 8}, 633--671.

\bibitem{Baerheim} Baerheim, R. (1993) Harmonic decomposition of the
  anisotropic elasticity tensor. {\it Quarterly J.\ Mech.\ Appl.\
    Math.} {\bf 46}, 391--418.

\bibitem{Bona1} B\'ona, A., Bucataru, I. \& Slawinski, M. A. (2004) Material symmetries of elasticity tensors, 
{\it  Quarterly J. Mech. Appl. Math. } {\bf 57}, 583--598.

\bibitem{Bona2}  B\'ona, A., Bucataru, I. \& Slawinski, M. A. (2007) Material symmetries versus wavefront
symmetries. {\it Quarterly J. Mech. Appl. Math. } {\bf  60}, 73--84.
\bibitem{Campanella} A. Campanella and M. L. Tonon (1994) {A note on the Cauchy relations,} {\it Meccanica} {\bf 29 }, 105-108. 

\bibitem{Cowin1989} Cowin, S.~C. (1989) Properties of the anisotropic
  elasticity tensor. {\it Quarterly J.\ Mech.\ Appl.\ Math.} {\bf
    42}, 249--266. Corrigenda  ibid. (1993) {\bf 46}, 541--542.

\bibitem{Cowin1992} Cowin, S.~C.\ \& Mehrabadi, M.~M. (1992) The structure of
  the linear anisotropic elastic symmetries, {\it J.\ Mech.\ Phys.\
    Solids} {\bf 40}, 1459--1471.

\bibitem{Cowin} S. C. Cowin and M. M. Mehrabadi (1995)  Anisotropic symmetries of linear elasticity, {\it Appl. Mech. Rev.,} {\bf 48}(5), 247--285.

\bibitem{Fedorov} Fedorov, F. I. (2013) {\it Theory of elastic waves in crystals,} Springer Science \& Business Media.

\bibitem{Haus} Hauss\"uhl, S. 2007 {\it Physical Properties of
    Crystals: An Introduction.} Weinheim, Germany: Wiley-VCH.

\bibitem{Cauchy} F.~W.~Hehl and Y.~Itin (2002) The Cauchy relations in
  linear elasticity theory,  {\it J.\ Elasticity} {\bf 66}, 185--192.



\bibitem{Itin-Hehl} Y.~Itin and F.~W.~Hehl (2013) The constitutive
  tensor of linear elasticity: its decompositions, Cauchy relations,
  null Lagrangians, and wave propagation, {\it J.\ Math.\ Phys.} {\bf  54},
  042903 (2013).

\bibitem{Itin-Hehl2} Y.~Itin and F.~W.~Hehl (2015)  Irreducible decompositions of the elasticity tensor under the linear and orthogonal groups and their physical consequences. {\it Journal of Physics: Conference Series} {\bf 597}(1) 012046.

\bibitem{Leibfried} G. Leibfried, (1955) Gittertheorie der mechanischen und thermischen Eigenschaften
der Kristalle. In {\it Handbuch der Physik}, Vol. VII/1, Kristallphysik I; S.
Fl¨ugge, ed., Springer, Berlin  pp.104-324.

\bibitem{Marsden} Marsden, J.~E.\ \& Hughes, T.~J.~R. (1983) {\it
    Mathematical Foundations of Elasticity}, Englewood Cliffs, NJ:
  Prentice-Hall.

\bibitem{Moakher-Norris} Moakher, M.,  Norris, A. N. (2006). The closest elastic tensor of arbitrary symmetry to an elasticity tensor of lower symmetry, {\it Journal of Elasticity}  {\bf 85} (3), 215-263.

\bibitem{Nayfeh} A. H. Nayfeh, {\it Wave propagation in layered anisotropic media: with applications to composites} (North-Holland, Amsterdam, 1995)

\bibitem{Norris} Norris, A. N. (2007)  Quadratic invariants of elastic moduli, {\it Quarterly J. Mech. Appl. Math.} {\bf 60} (3), 367--389.

\bibitem{Norris1}  Norris, A.N., (2006) Elastic
 moduli approximation of higher symmetry for the acoustical properties of an anisotropic  material, {\it Journal of Acoustical Society of America }
{\bf 119} (4), 2114-2121


\bibitem{Podio-Guidugli} P. Podio-Guidugli, (2000) {\it A Primer in Elasticity}, Kluwer, Dordrecht.

\bibitem{SokolElast} Sokolnikoff, I.~S. (1956) {\it Mathematical Theory
    of Elasticity,} 2nd edn. New York: McGraw-Hill.

\bibitem{Surrel} Y. Surrel, { A new description of the tensors of elasticity based upon irreducible representations}, Eur. J. Mech. A/Solids 12 (1993)  219--235.  

\bibitem{Ting} T.~C.~T.~Ting, (1987) Invariants of anisotropic elastic constants, {\it Quarterly J. Mech. Appl. Math.} {\bf 40}  431--448.

\bibitem{Weiner} J.H. Weiner, {\it Statistical Mechanics of Elasticity} (Dover, Mineola, New York, 2002)

\bibitem{Weyl} Weyl, H. (1997) {\it The classical groups: their invariants and representations} Vol. 1, Princeton university press.

\bibitem{Xiao} H. Xiao, (1998) { On anisotropic invariants of a symmetric tensor: crystal classes, quasi-crystal classes and others,} {\it Proc. R. Soc.} London A  {\bf 454} 1217--1240. 
\end{thebibliography}
\end{document}